\documentclass[twocolumn,aps,prl,reprint]{revtex4-1}
\usepackage{amsmath}
\usepackage{xcolor}

\usepackage{braket}
\usepackage{amsfonts}
\newcommand{\Tr}{\text{Tr}}
\usepackage{verbatim}
\usepackage{subcaption}
\usepackage{graphicx}
\usepackage[margin=0.75in]{geometry}

\begin{document}

\title{Anti-Hawking Phenomena around a Rotating BTZ Black Hole}
\author{Matthew P. G. Robbins${}^{1,2,3}$ and Robert B. Mann${}^{1,2,3,4}$}
\affiliation{Department of Physics and Astronomy, University of Waterloo, Waterloo ON, Canada, N2L 3G}
\affiliation{Perimeter Institute for Theoretical Physics, 31 Caroline Street North, Waterloo ON, Canada, N2L 2Y5}
\affiliation{Waterloo Centre for Astrophysics, University of Waterloo, Waterloo ON, Canada, N2L 3G1}
\affiliation{Institute for Quantum Computing, University of Waterloo, Waterloo ON, Canada, N2L 3G}

\begin{abstract}
 In both flat and curved spacetimes, there are weak and strong versions of the anti-Unruh/anti-Hawking effects, 
 in which the KMS field temperature is anti-correlated with the response of a detector and its
 inferred temperature.  We investigate for the first time the effects on the weak and strong anti-Hawking effects for an Unruh-DeWitt detector orbiting a BTZ black hole in the co-rotating frame.    
 We find that rotation can significantly amplify  the strength of the weak anti-Hawking effect, whereas it can either amplify or reduce the strength of the strong anti-Hawking effect depending on boundary conditions.   For the strong anti-Hawking effect, we find a non-monotonic relationship between the angular momentum and detector temperature for each boundary condition. In addition, we note that the weak anti-Hawking effect is independent of a changing AdS length, while a longer AdS length increases the temperature range of the strong anti-Hawking effect.
 \end{abstract}

\maketitle

\section{Introduction}
The nature of the quantum vacuum is fundamental to the study of quantum information. By coupling the vacuum field to particle detectors, it is possible to gain understanding of the quantum vacuum and its response to the structure of spacetime. This
is particularly interesting if topological features or horizons are present.  
These detectors are often taken to be simple two-level quantum systems (known as Unruh-DeWitt detectors) that interact with an underlying scalar field, a model that captures the essential features of the light-matter interaction \cite{MartinMartinez:2012th,Funai:2021jpc}.  Such detectors have been employed to study the structure of spacetime \cite{Smith2016,Ng2017}, black holes \cite{Henderson2018,Tjoa2020}, and the thermality of de Sitter spacetime \cite{Steeg2009,Huang2017}.

Consider a detector with uniform acceleration in flat spacetime. The detector will experience the Unruh effect, in which the temperature increases in proportion to acceleration \cite{Fulling1973,Davies1975,Unruh1976}, and  heat up.  This effect arises because the temperature of the vacuum of one set of modes is different than the vacuum temperature of a second set of modes. 
Though highly idealized in its original assumptions (such as that of an eternally uniformly accelerating detector), a model-independent derivation of the Unruh effect has been given  in the context of axiomatic quantum field theory \cite{sewell1982quantum}, with the field temperature (given by the Kubo-Martin-Schwinger (KMS) condition \cite{Kubo1957,Martin1959,Haag1967}) and the temperature measured by the detector being the same. There have since been many demonstrations that detectors undergoing other forms of acceleration (non-uniform, circular) get hot \cite{Bell:1986ir,Costa:1987vv,Percoco:1991iv,Villalba:1999uj,Ostapchuk:2011ud,Hu:2012jr,Doukas2013}. 
 In these more general situations the field temperature is  positively correlated with the detector temperature, with the latter a monotonically increasing function of the former.

In the last few years it has been shown that some physical situations exhibit the so-called \textit{anti}-Unruh effect instead, in which  the temperature of the field is no longer  positively correlated with that measured by the detector \cite{Brenna2016,Garay2016}. 
 This anti-Unruh effect can be split into two cases: a  weak Anti-Unruh effect (as the temperature of the field increases, the detector clicks less often) and a strong Anti-Unruh effect (the field temperature and detector temperature are inversely related) \cite{Garay2016}. 

When considering black holes, the Hawking effect is the analogue of the Unruh effect \cite{Birrell1982}. 
While in general detector temperatures are positively correlated with the field temperature outside a black hole
\cite{Hodgkinson2012,Smith2014,Hodgkinson:2014iua,Ng:2014kha}, recently an anti-Hawking effect was shown to also exist, in which a static Unruh-DeWitt (UdW) detector exhibited both strong and weak versions of the phenomenon \cite{Henderson:2019uqo}.  This was explicitly demonstrated for  the (2+1) dimensional static Banados-Teitelboim-Zanelli (BTZ) black hole.  
For sufficiently small black holes, the temperature measured by the detector would decrease as the Kubo-Martin Schwinger (KMS) field temperature of the Hawking radiation increased. The anti-Hawking effect has since been observed  for a 
broader range of boundary conditions \cite{Campos:2020twd}, though its weak version is not observed for massless topological black holes in four spacetime dimensions \cite{Campos:2020lpt}.

The effects of spacetime dragging due to rotation on these phenomena are much less understood.
The quantum vacuum around a rotating black hole
is known to exhibit features significantly different from  its non-rotating counterpart \cite{Winstanley_2001,Casals_2008,Ottewill:2000yr}.  Several investigations of the behaviour of quantum scalar fields in
the background of a rotating BTZ black hole have been carried out \cite{Mann:1996ze,Singh:2011gd,Bussola:2017wki,Meitei:2018mgo}, and studies   investigating the response of Unruh-DeWitt detectors   in such spacetimes have also been undertaken \cite{Hodgkinson2012}.  Recently it was shown
that rotation has very significant effects on the entanglement harvesting abilities of   UdW  detectors, with
the harvested entanglement being considerably amplified at intermediate distances (about 20-50 horizon radii) from the
black hole \cite{Robbins2020}.

Motivated by this, we study here  the implications of rotation for the anti-Hawking effect.   We find that  rotation  increases the intensity of the weak anti-Hawking effect, but has   a negligible influence on its threshold critical temperature. However for the strong anti-Hawking effect,  we find that there is a strong dependence on the angular momenta, with the effect becoming stronger or weaker depending on the boundary conditions.  The  influence of AdS length on the strong and weak versions of the effect is likewise distinct:  the weak anti-Hawking effect is independent of AdS length whereas the strong version sees an increased temperature range.

\section{Unruh-DeWitt Detectors}

To model the interaction between the detectors and the field, we take the detectors to be two-level quantum systems with ground state $\ket{0}_D$ and excited state $\ket{1}_D$, separated by an energy gap $\Omega_D$. We shall assume that these detectors have a spacetime trajectory $x_D(\tau)$. The interaction Hamiltonian is
\begin{align}
H_D=\lambda\chi_D(\tau)\left(e^{i\Omega_D\tau}\sigma^+ + e^{-i\Omega_D\tau}\sigma^-\right)\otimes\phi[x_D(\tau)]\,,
\end{align}
where the switching function dictating the duration of the interaction between the detector and field is $\chi_D(\tau)$. $\lambda\ll1$ is the field-detector coupling constant, and the ladder operators that raise and lower the energy levels of the detectors are $\sigma^+= \ket{1}_D\bra{0}_D$, $\sigma^-= \ket{0}_D\bra{1}_D$, respectively.

Let the initial state of the detector-field system be $\ket{\Psi_i}=\ket{0}_D\ket{0}$. The final state after a time $t$ is then $\ket{\Psi_f}=U(t,0)\ket{\psi_i}$, where $U(t,0)=\mathcal{T}e^{-i\int dt\left[\frac{d\tau_D}{dt}H_D(\tau_D))\right]}$, with $\mathcal{T}$ being the time-ordering operator. With the reduced density operator   $\rho=\Tr_\phi\ket{\Psi_f}\bra{\Psi_f}$ being the state of the system after integrating over the field's degrees of freedom, we have \cite{Smith2016,Smith2017}
\begin{align}
\rho_{AB}=\begin{pmatrix}
1-P_D &0\\
0&P_D
\end{pmatrix}
+\mathcal{O}(\lambda^4)\,,
\end{align}
where
\begin{widetext}
\begin{align}
P_D=\lambda^2\int d\tau_Dd\tau_D'\chi_D(\tau_D)\chi_D(\tau_D')e^{-i\Omega_D(\tau_D-\tau_D')}W(x_D(\tau_D),x_D(\tau_D'))
\label{eq: PD}
\end{align}
\end{widetext}
 is the detector's transition probability. We note that this quantity depends on the two-point correlation function, $W(x,x')=\braket{0|\phi(x)\phi(x')|0}$ (also called the Wightman function) of the vacuum. From the transition probability, we can then define the response function,
 \begin{align}
 \mathcal{F}=\frac{P_D}{\lambda^2\sigma}\,,
 \end{align}
 where $\sigma$ describes the timescale of interaction between the field and the detector. In this paper, we shall focus on a Gaussian switching function, $\chi_D(\tau)=e^{-\frac{\tau^2}{2\sigma^2}}$.

 We will  consider fields whose Wightman functions   obey the relation
 \begin{equation}\label{KMScond}
 W(\tau-i/T_{KMS},\tau')=W(\tau',\tau)
 \end{equation}
 known as the Kubo-Martin-Schwinger (KMS) condition  \cite{Kubo1957,Martin1959,Haag1967}.  The quantity $T_{KMS}$ 
 in \eqref{KMScond} can be regarded as the temperature of the quantum field in the spacetime.  It depends only on
 the nature of the quantum field and the spacetime background.
 
Correspondingly, we can also define the detector's temperature from its excitation to de-excitation ratio (EDR). Let
 \begin{align}
 \mathcal{R}=\frac{\mathcal{F}(\Omega)}{\mathcal{F}(-\Omega)}\,,
 \end{align}
 such that there exists a temperature, $T$, that obeys the same form of the KMS condition \cite{Fewster2016},
 \begin{align}
  \mathcal{R}=e^{-\Omega/T}\,.
 \end{align}
Labelling the temperature that obeys this condition by $T_{EDR}$, we have
\begin{align}
T_{EDR}=-\frac{\Omega}{\log\mathcal{R}}
\end{align}
The quantity $T_{EDR}$ can be regarded as the temperature that the UdW detector registers in the spacetime.

Normally we expect $T_{EDR}$ and $T_{KMS}$ to be positively correlated: as the black hole gets hotter, the field temperature  increases
and the temperature registered by the UdW detector likewise increases.   This is indeed the case for most situations 
in black hole physics. As noted above, it was recently shown that this is not always the case  \cite{Henderson:2019uqo}, and
that sometimes the contrary situation, known as the anti-Hawking effect, occurs.  As with the anti-Unruh effect \cite{Brenna2016,Garay2016}, we define  
\begin{align}\label{weakAH}
&\frac{d\mathcal{F}(\Omega)}{dT_{KMS}}<0  \quad \textrm{weak}\\
&  \frac{\partial T_{EDR}}{\partial T_{KMS}} < 0
\quad  
\textrm{strong} 
\label{strongAH}
\end{align} 
for the weak and strong anti-Hawking effects respectively. 
 
 \section{Rotating BTZ Black holes}
 
 We can write the action of our system as $S=S_{EH}+S_{\phi}$, where 
 \begin{align}
 S_{EH}=\frac{1}{16\pi}\int R\sqrt{-g}d^3x
 \end{align}
 is the Einstein-Hilbert action ($R$ is the Ricci scalar and $g$ is the determinant of the metric tensor $g_{\mu\nu}$) and
 \begin{align}
 S_{\phi}=-\int\left(\frac{1}{2}g^{\mu\nu}\partial_{\mu}\phi\partial_{\nu}\phi+\frac{1}{16}R\phi^2\right)\sqrt{-g}d^3x
 \end{align}
 is the action for a conformally-coupled scalar field $\phi$. We are interested in analyzing both the KMS temperature of the field and the EDR temperature of a detector near a rotating BTZ black hole, whose line element is  \cite{Banados1992}
\begin{equation}\label{btzmet}
ds^2=-(N^\perp)^2dt^2+f^{-2}dr^2+r^2(d\phi+N^{\phi}dt)^2
\end{equation}
where, $N^\perp=f=\sqrt{-M+\frac{r^2}{\ell^2}+\frac{J^2}{4r^2}}$ and $N^\phi =-\frac{J}{2r^2}$ with $M=\frac{r_+^2+r_-^2}{\ell^2}$ the mass of the black hole  and $J=\frac{2r_+ r_-}{\ell}$ its angular momentum. The Hawking temperature is 
\begin{align}
T_H=\frac{1}{2\pi\ell}\left(\frac{r_+^2-r_-^2}{r_+}\right)\,,
\end{align}
where the inner and outer horizon radii are denoted by $r_-$ and $r_+$, and $\ell$ is the AdS length. Note that $|J|\leq M\ell$, with extremality occurring when $r_+=r_-$ (i.e. $J=M\ell$).
\begin{figure*}[t!]
\centering
\begin{minipage}{\textwidth}
  \centering
    \includegraphics[width=0.6\textwidth]{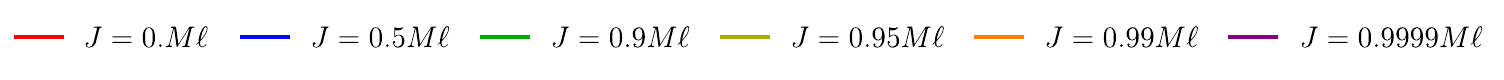}
\end{minipage}\\
\begin{minipage}{0.3\textwidth}
  \centering
    \includegraphics[width=\textwidth]{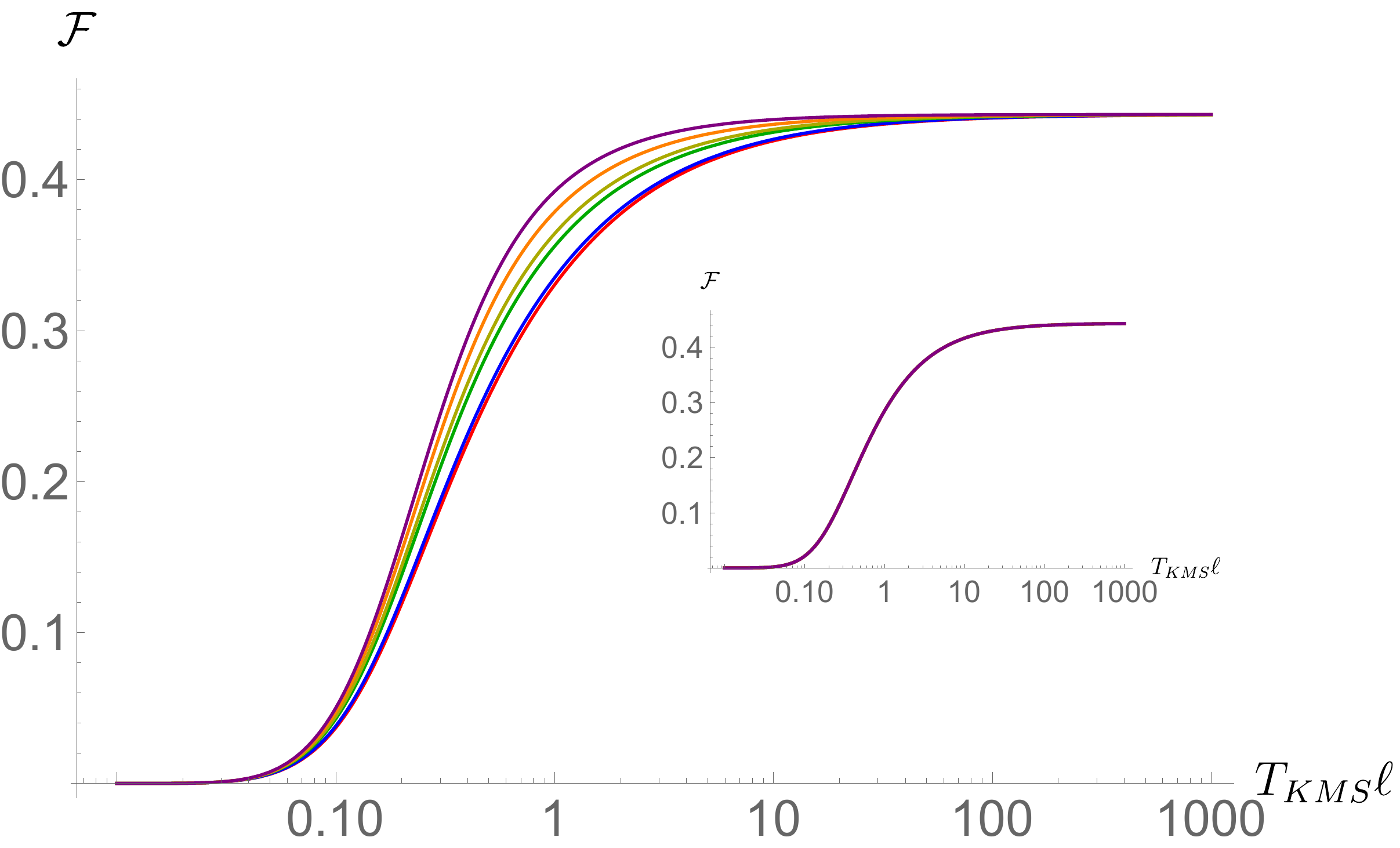}
    \caption*{(a) $\zeta=1$}
\end{minipage}
\begin{minipage}{0.3\textwidth}
  \centering
    \includegraphics[width=\textwidth]{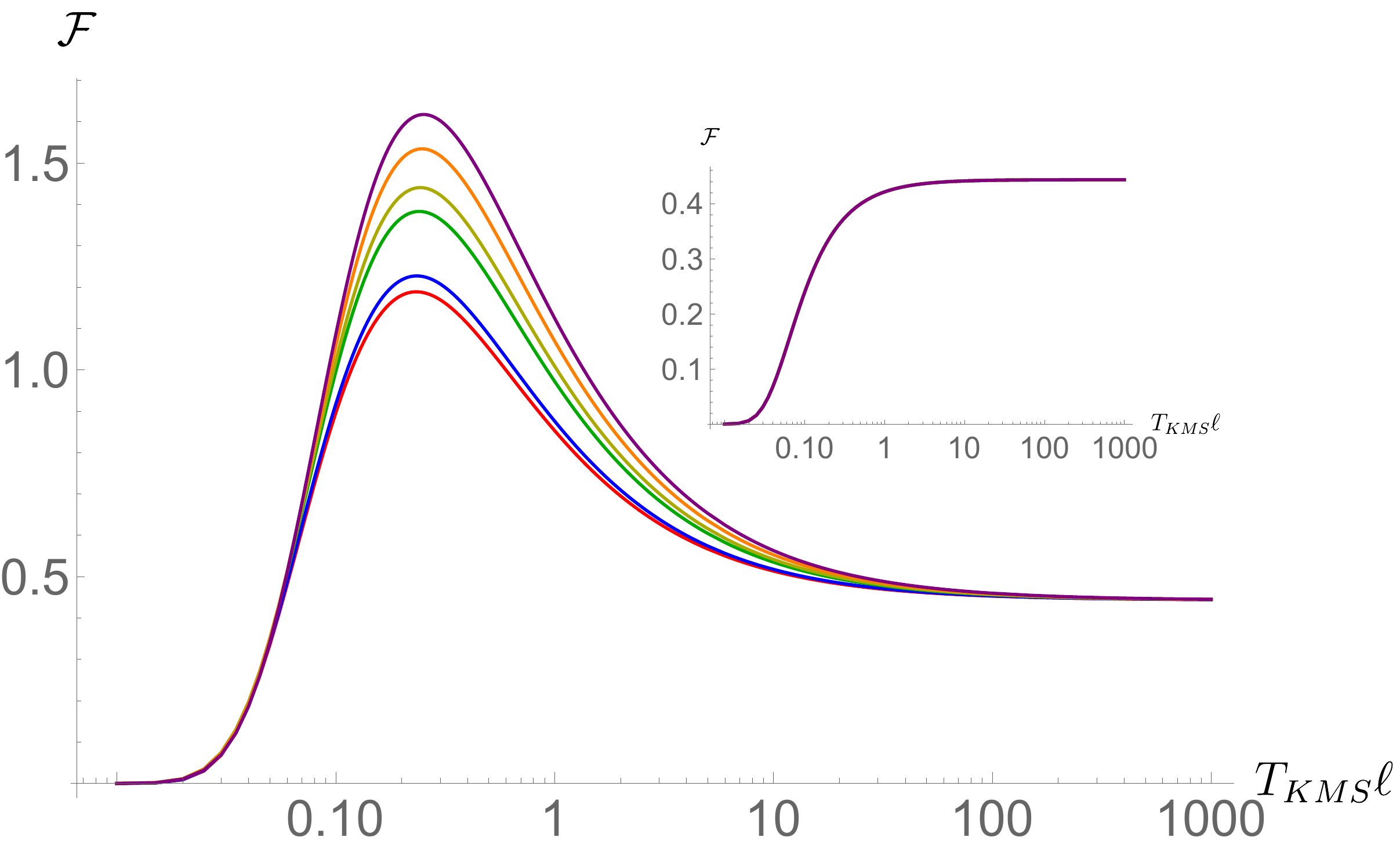}
    \caption*{(b) $\zeta=0$}
\end{minipage}
\begin{minipage}{0.3\textwidth}
  \centering
  \centering
    \includegraphics[width=\textwidth]{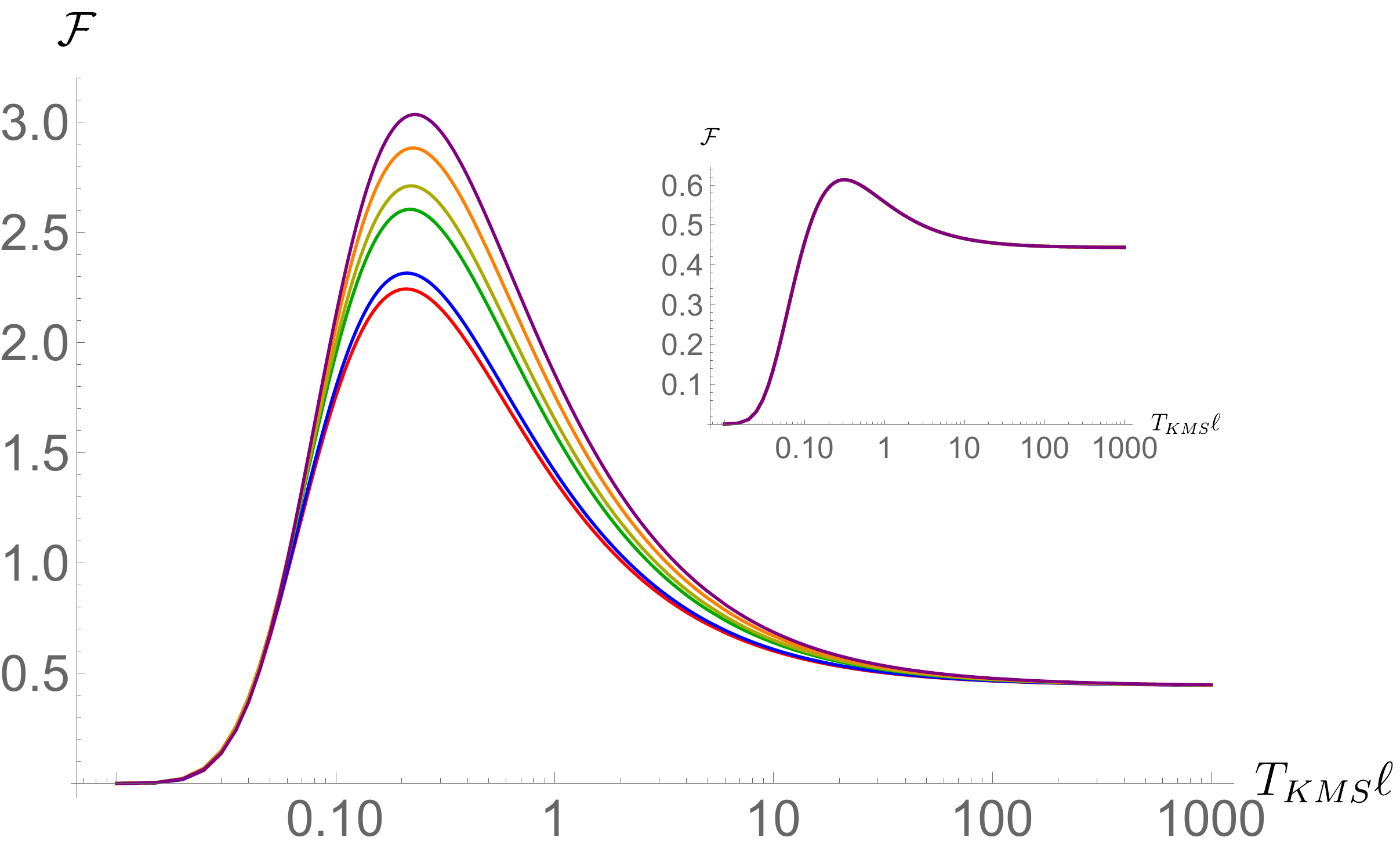}
    \caption*{(c) $\zeta=-1$}
\end{minipage}
\caption{Response functions for a black hole of mass $M=1/10$ for Dirichlet,  transparent, and Neumann boundary conditions and an energy gap of $\Omega\ell=1/10$. The inset plots correspond to $M=100$. As expected, the rotation of the black hole has a smaller effect for larger masses. As the mass of the black hole increases, the weak anti-Hawking effect goes away for $\zeta=1$ and $\zeta=0$.
Note that for $\zeta=-1$, the weak anti-Hawking effect is still present even for large mass black holes, with the distinctions
between the different rotation parameters  so tiny that the curves effectively all overlap.
}
\label{fig: FJ_LargerMass}
\end{figure*}

\begin{figure*}[t!]
\centering
\begin{minipage}{\textwidth}
  \centering
    \includegraphics[width=0.6\textwidth]{PDLegendConstantM.pdf}
\end{minipage}\\
\begin{minipage}{0.3\textwidth}
  \centering
    \includegraphics[width=\textwidth]{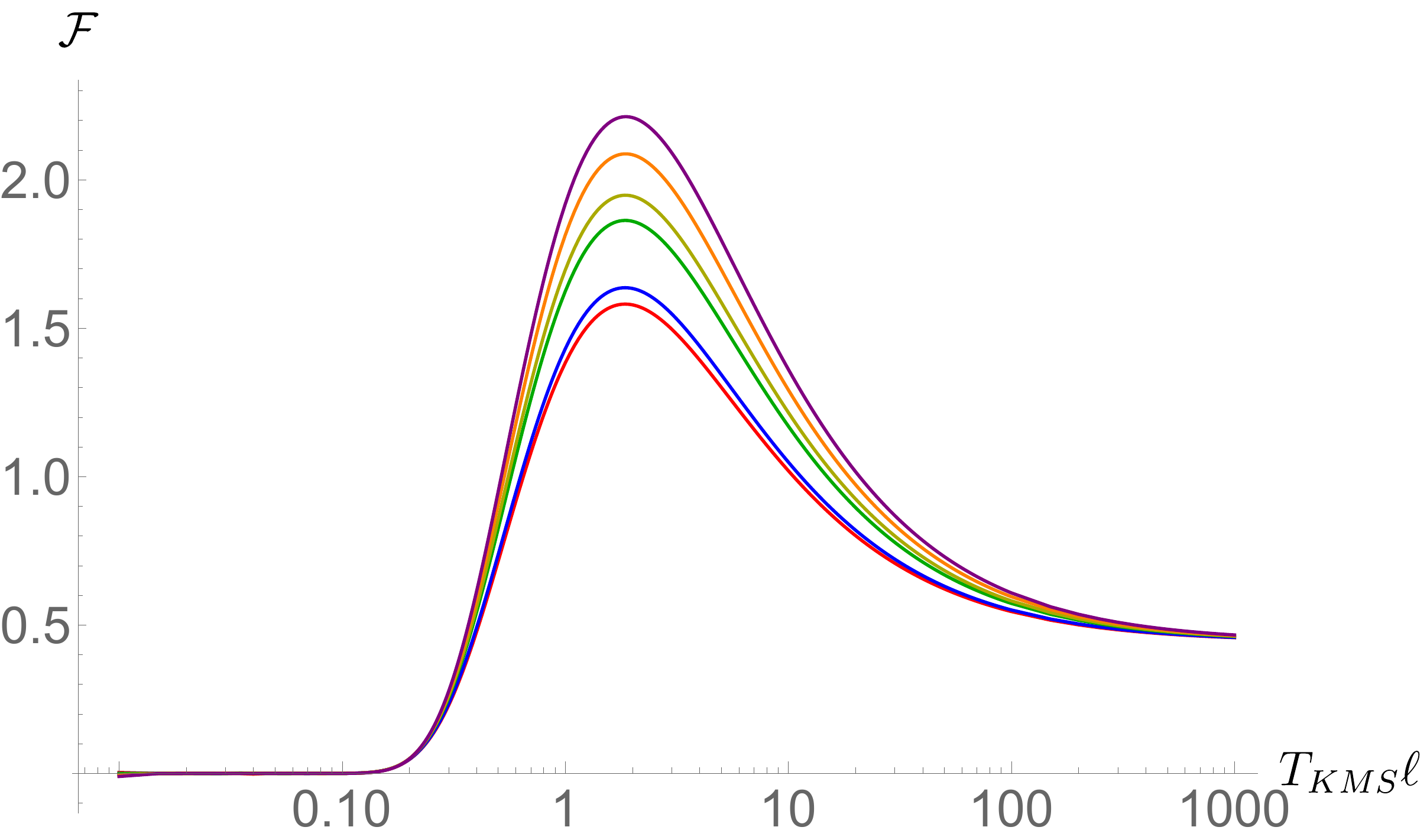}
    \caption*{(a) $\Omega\ell=1$}
\end{minipage}
\begin{minipage}{0.3\textwidth}
  \centering
    \includegraphics[width=\textwidth]{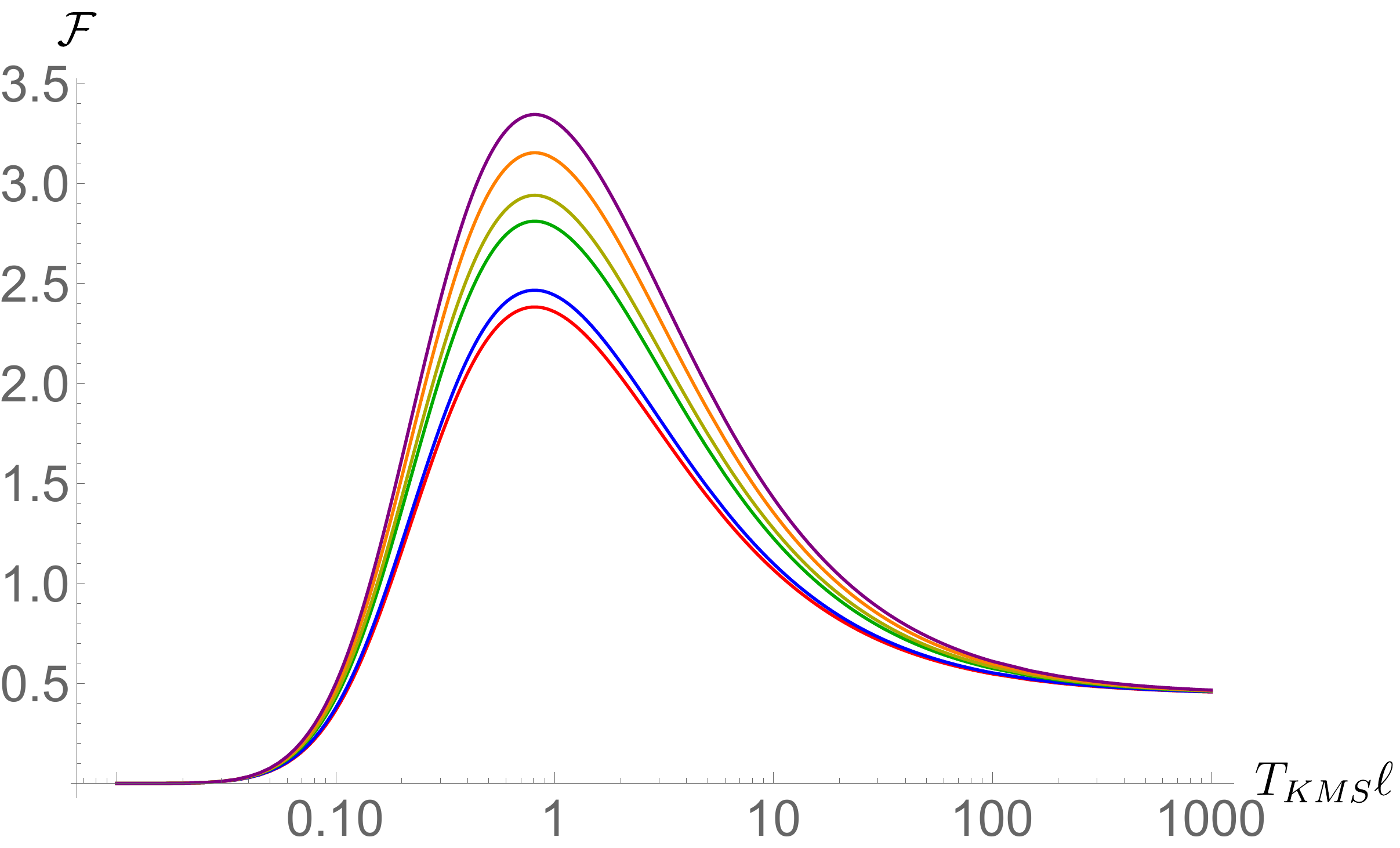}
    \caption*{(b) $\Omega\ell=1/10$}
\end{minipage}
\begin{minipage}{0.3\textwidth}
  \centering
  \centering
    \includegraphics[width=\textwidth]{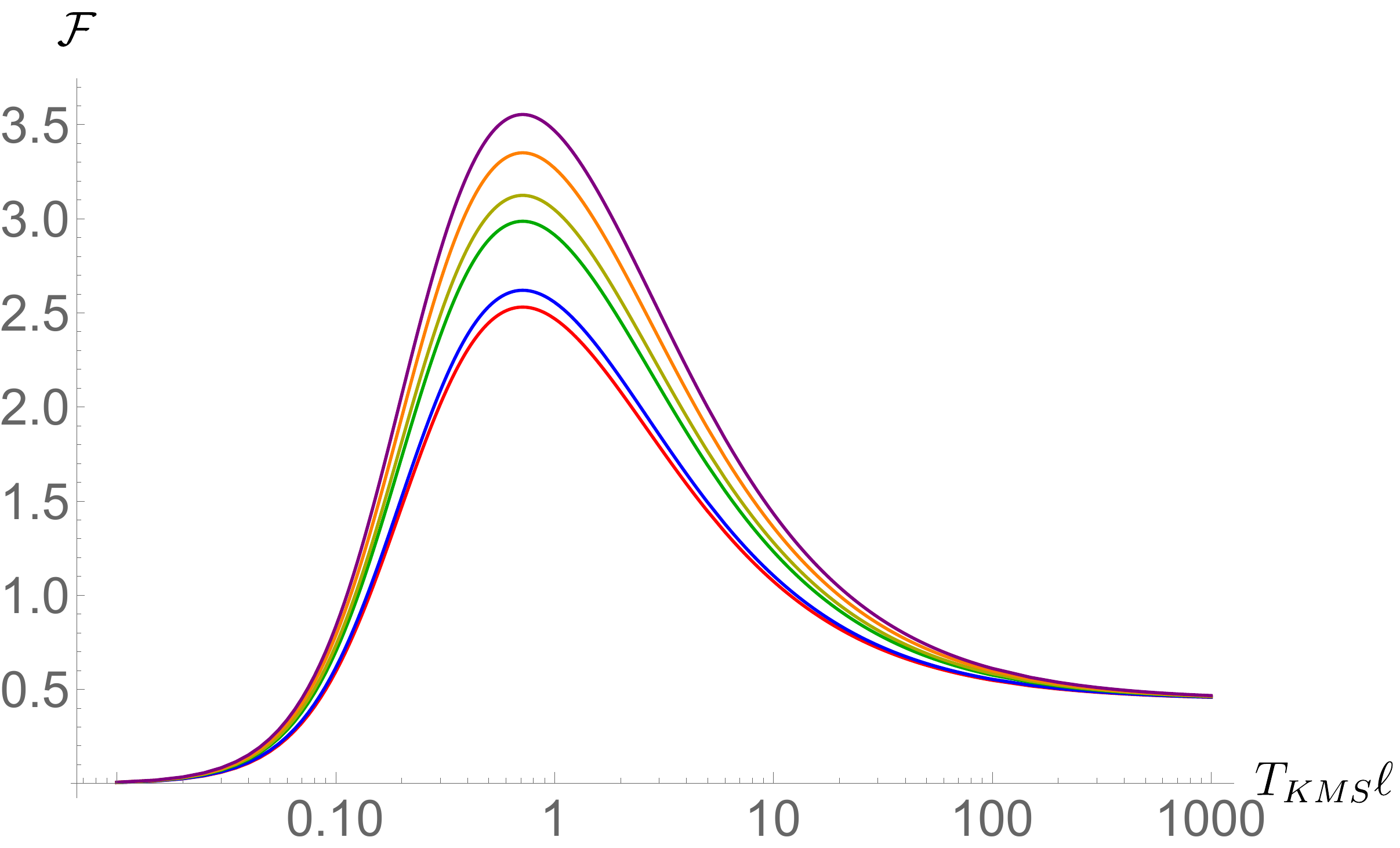}
    \caption*{(c) $\Omega\ell=1/100$}
\end{minipage}
\caption{Response of a rotating BTZ black hole with mass $M=1/1000$ and Dirichlet boundary conditions ($\zeta=1$). We note that for transparent and Neumann boundary conditions, qualitatively similar results are obtained.
}
\label{fig: FJ}
\end{figure*}

In the Hartle-Hawking vacuum, a conformally coupled-scalar field has a Wightman function that can be written as the image sum over the Wightman functions for AdS${}_3$ \cite{Lifschytz1994,Carlip1998},
\begin{align}
W_{BTZ}(x,x')=\sum_{n=-\infty}^\infty \eta^nW_{AdS_3}(x,\Gamma^nx')
\end{align}
where $\Gamma x'$ takes $(t,r,\phi)\to(t,r,\phi+2\pi)$, $\eta=1$ is an untwisted scalar field and $\eta=-1$ is a twisted scalar field. This yields \cite{Hodgkinson2012,Smith2014}
\begin{widetext}
\begin{equation}
\begin{aligned}
W_{BTZ}&=\frac{1}{4\pi}\frac{1}{2\sqrt{\ell}}\sum_{n=-\infty}^{n=\infty}\eta^n\left(\frac{1}{\sqrt{\sigma_{\epsilon}(x,\Gamma^nx')}} -\frac{\zeta}{\sqrt{\sigma_{\epsilon}(x,\Gamma^nx')+2}}\right)
\label{eq: sum}
\end{aligned}
\end{equation}
where
\begin{equation}
\begin{aligned}
\sigma_\epsilon(x,\Gamma^nx')^2=&-1+\sqrt{\alpha(r)\alpha(r')}\cosh\left[\frac{r_+}{\ell}(\Delta\phi-2\pi n)-\frac{r_-}{\ell^2}(t-t')\right]\\
&-\sqrt{(\alpha(r)-1)(\alpha(r')-1)}\cosh\left[\frac{r_+}{\ell^2}(t-t')-\frac{r_-}{\ell}(\Delta\phi-2\pi n)\right]
\label{eq: sigma 0}
\end{aligned}
\end{equation}
and 
\begin{align}
\alpha(r)&=\frac{r^2-r_-^2}{r_+^2-r_-^2}\qquad
\Delta\phi=\phi-\phi' 
\end{align}
\end{widetext}
with the respective boundary conditions as $\zeta=1$ (Dirichlet), $\zeta=0$ (transparent), and $\zeta=-1$ (Neumann). We shall take the detector to have switching function $\chi_D(\tau_D)=e^{-\tau_D^2/2\sigma^2}$ and only consider untwisted scalar fields with $\eta=1$.

To calculate the transition probabilities, we will work in the co-rotating frame of the detectors  \cite{Hodgkinson2012}: \begin{align}
t&=\frac{\ell r_+\tau}{\sqrt{r^2-r_+^2}\sqrt{r_+^2-r_-^2}} \label{eq: CRM t}\\
\phi&=\frac{r_-\tau}{\sqrt{r^2-r_+^2}\sqrt{r_+^2-r_-^2}} \label{eq: CRM phi}\ .
\end{align}
in which case \cite{Vagenas2002}
\begin{align}\label{temprel}
T_{KMS}=T_{H}/\gamma\,,
\end{align}
where
\begin{align}\label{lorfac}
\gamma=\frac{\sqrt{r^2-r_+^2}\sqrt{r_+^2-r_-^2}}{r_+}
\end{align}
is the Lorentz factor.
Straightforward calculations show that we can rewrite equation \eqref{eq: PD} as $P_D=\sum_{n=-\infty}^\infty\eta^n\left\{I_n^--\zeta I_n^+\right\}$, where
\begin{align}
I_n^\pm =K_P\int_{-\infty}^{\infty}dz\frac{e^{-a\left(z-\frac{2\pi nr_-}{\ell}\right)^2}e^{-i\beta\left(z-\frac{2\pi nr_-}{\ell}\right)}}{\sqrt{\left(\cosh(\alpha_n^\pm)-\cosh\left(z\right)\right)}}
\label{eq: In}
\end{align}
\begin{widetext}
and
\begin{align}
K_P&=\frac{\lambda ^2 \sigma_D}{4 \sqrt{2 \pi }}\\
a&=\frac{1}{(4\pi T_{KMS}\sigma_D)^2} \qquad 
\beta =\frac{\Omega_D}{2\pi T_{KMS}}\\
\cosh(\alpha_n^\pm)&= \pm4\ell^2\pi^2T_{KMS}^2+(1+4\ell^2\pi^2T_{KMS}^2)\cosh\frac{2\pi n r_+}{\ell}
\end{align}

In the limit of an infinite interaction time (i.e. $\sigma\to\infty$), we note that we can write the $n=0$ term (corresponding to AdS spacetime) analytically as
\begin{align}
\lim_{\sigma\to\infty}P_{D,n=0}=\lim_{\sigma\to\infty}I_0=\frac{\sqrt{\pi}}{4}\left[1-\tanh\frac{\Omega_D}{2T_{KMS}}\right]\left[1-\zeta P_{-1/2+i\beta}\left(\cosh\alpha_0^+\right)\right]
\label{eq: I0}
\end{align}
\end{widetext}

To investigate the influence of rotation (and AdS length) on the weak and strong anti-Hawking effects, we must determine the dependence of the response function and EDR temperature on the KMS temperature.  To vary the latter, we locate the detector
at   $r=R_D$ in the co-rotating frame and solve \eqref{temprel} and \eqref{lorfac} for $R_D$ in terms of $T_{KMS}$ and the
other parameters.  The response $P_D$ in \eqref{eq: PD} and the EDR temperature  are then  functions of $T_{KMS}$.

\section{Weak Anti-Hawking Effect for Rotating BTZ Black Holes}

 To set the context for our investigation, we first compare the situation for large and intermediate mass black holes with a detector energy gap of $\Omega_D\ell=1/10$ (with other energy gaps yielding qualitatively similar results). This is shown in   Figure~\ref{fig: FJ_LargerMass}, where we plot detector response as a function of $T_{KMS}$.  The general trend is that for both masses and all boundary conditions the response is suppressed at small KMS temperatures  and asymptotes to a constant value at a large ones. Apart from these two general features, there is notably distinct behaviour as these parameters are varied.
In the main figure, we depict the situation for an intermediate mass  $M=1/10$, where we see that rotation has marginal impact at small and large KMS temperatures, but  significantly amplifies  detector response at  intermediate KMS temperatures.  An even larger influence is due to boundary conditions,
where we observe that the weak anti-Hawking effect is absent for Dirichlet boundary conditions, but present for the other two. The
negative slope to the right of the peak is steeper for Neumann boundary conditions, indicative of increasing strength of the weak effect as $\zeta$ decreases.
In the inset of each subfigure, we consider the large mass $M=100$ case, where we see that the weak anti-Hawking effect disappears for $\zeta=1,0$ boundary conditions, yet remains for $\zeta=-1$, recovering earlier results
 \cite{Henderson:2019uqo} for large-mass black holes.  We see that there is negligible dependence of the response on angular momentum in this large mass case for all KMS temperatures.

\begin{figure*}[t!]
\centering
\begin{minipage}{\textwidth}
  \centering
    \includegraphics[width=0.6\textwidth]{PDLegendConstantM.pdf}
\end{minipage}\\
\begin{minipage}{0.3\textwidth}
  \centering
    \includegraphics[width=\textwidth]{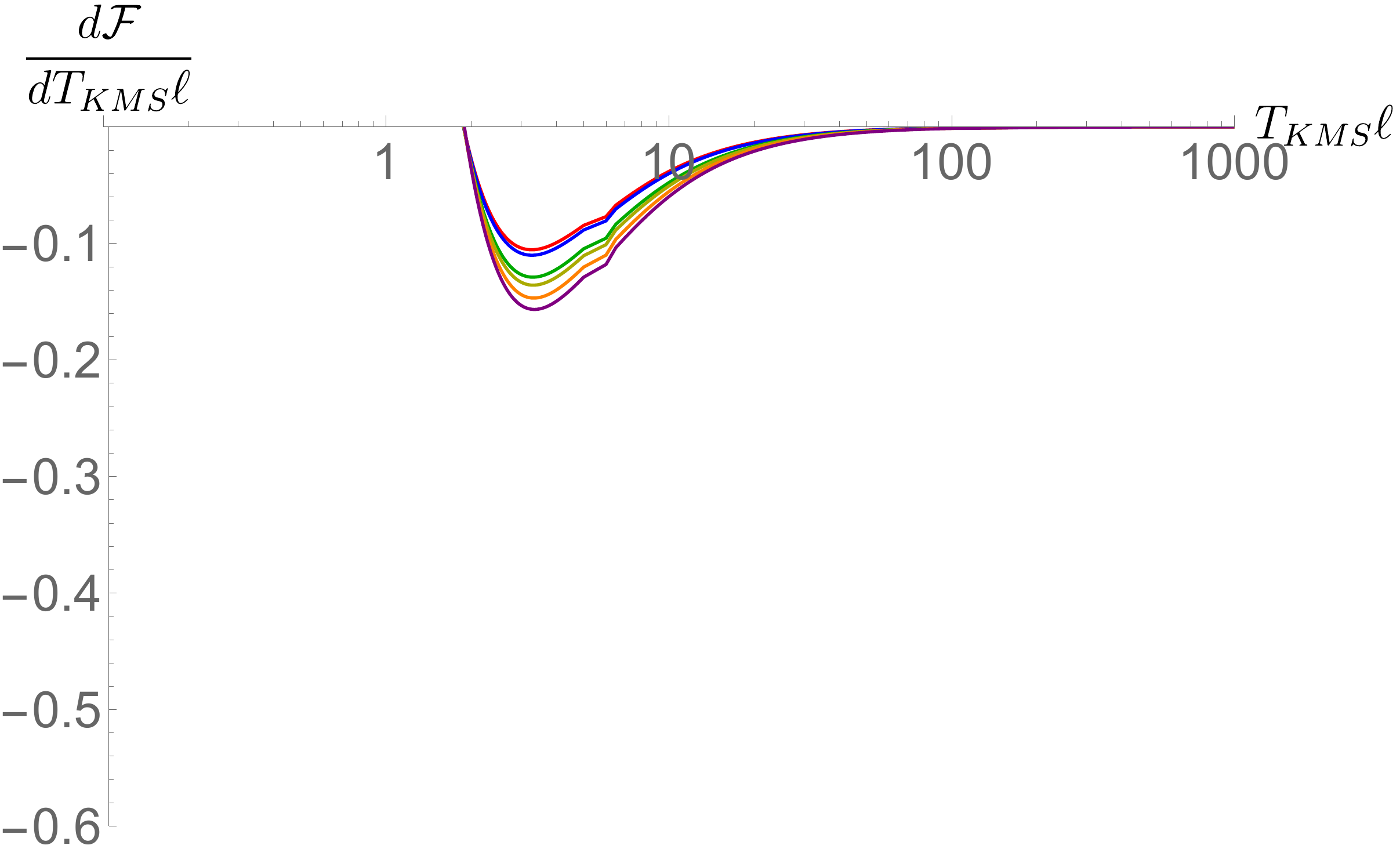}
    \caption*{(a) $\Omega\ell=1$}
\end{minipage}
\begin{minipage}{0.3\textwidth}
  \centering
    \includegraphics[width=\textwidth]{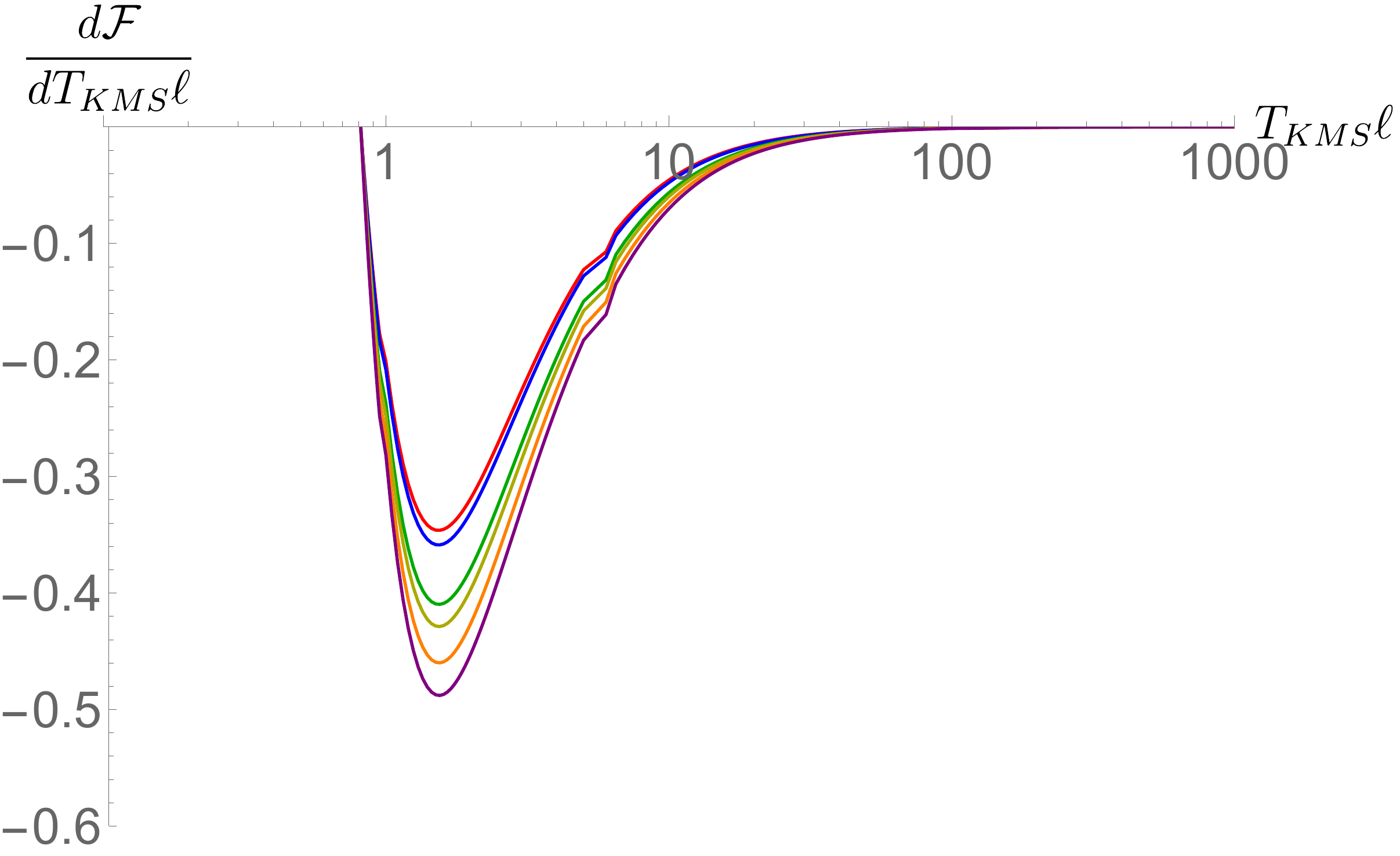}
    \caption*{(b) $\Omega\ell=1/10$}
\end{minipage}
\begin{minipage}{0.3\textwidth}
  \centering
  \centering
    \includegraphics[width=\textwidth]{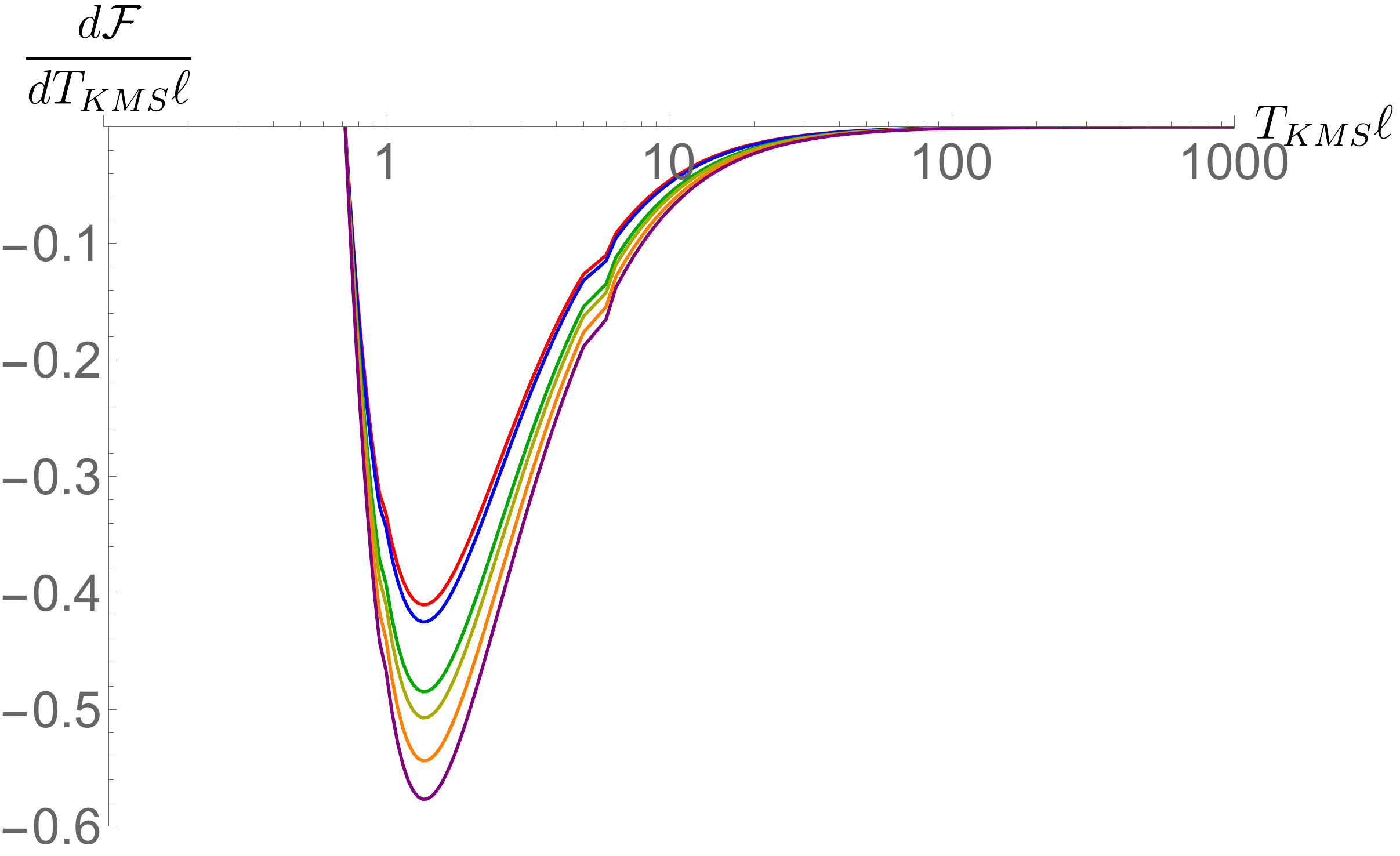}
    \caption*{(c) $\Omega\ell=1/100$}
\end{minipage}
\caption{ Derivative of the response with respect to the KMS temperature \eqref{weakAH} of a rotating BTZ black hole with mass $M=1/1000$, and Dirichlet boundary conditions ($\zeta=1$). We note that for transparent and Neumann boundary conditions, qualitatively similar results are obtained.}
\label{fig: FJ derivative}
\end{figure*}
 It is clear from this that rotational effects are more pronounced for smaller mass black holes, and so in Figure~\ref{fig: FJ}, we plot the dependence of the response function on  KMS temperature for  $M=1/1000$.   The weak effect is now clearly evident for  Dirichlet boundary conditions, with the peak response enhanced as much as sevenfold, compared to the left diagram in  Figure~\ref{fig: FJ_LargerMass}.  Similar results hold for the other boundary conditions.   As before, as rotation increases, the response is amplified for all values of $T_{KMS}$.
For all values of rotation and all gaps, the response asymptotes to a value of $P_D=\frac{\sqrt\pi}{4}$, in accord with equation \eqref{eq: I0}, noting from equation  \eqref{eq: In}  that 
  $I_n\to0$ for $n\neq0$. 

 The strength of the weak anti-Hawking effect depends on the magnitude of the 
 negative value of the slope of equation \eqref{weakAH}  after the peak.  We see that this
increases with decreasing gap, showing that smaller gap enhances the weak anti-Hawking effect, which we illustrate for Dirichlet boundary conditions in Figure \ref{fig: FJ derivative}.  The slope peaks at $\frac{d\mathcal{F}}{d T_{KMS}\ell}\approx-0.15$ for the large energy gap and $\frac{d\mathcal{F}}{d T_{KMS}\ell}\approx-0.6$ for the small gap. For each gap, we also see that the weak anti-Hawking effect is amplified with increasing rotation, by
as much as 50\% for near-extremal black holes,  for all gaps in the figure.
  
Furthermore, we find that the weak anti-Hawking effect occurs after a critical value of $T_{KMS}$   that depends on the detector's energy gap, but not on the rotation of the black hole, again evident from  Figure \ref{fig: FJ derivative}.  Though we have only
illustrated results for  Dirichlet boundary conditions $\zeta=1$, we emphasize that this critical value depends on $\zeta$,  with the critical temperature becoming smaller as $\zeta\to\-1$. 
 
Finally, we note that changing the AdS length will not change the strength of the weak effect. Physically, this is because the AdS length is the only length scale present (as $\sigma\to\infty$), and everything is calibrated against this length.

\section{Strong Anti-Hawking Effect for Rotating BTZ Black Holes}

\begin{figure*}[t!]
\centering
\begin{minipage}{\textwidth}
  \centering
    \includegraphics[width=0.6\textwidth]{PDLegendConstantM.pdf}
\end{minipage}\\
\begin{minipage}{0.3\textwidth}
  \centering
    \includegraphics[width=\textwidth]{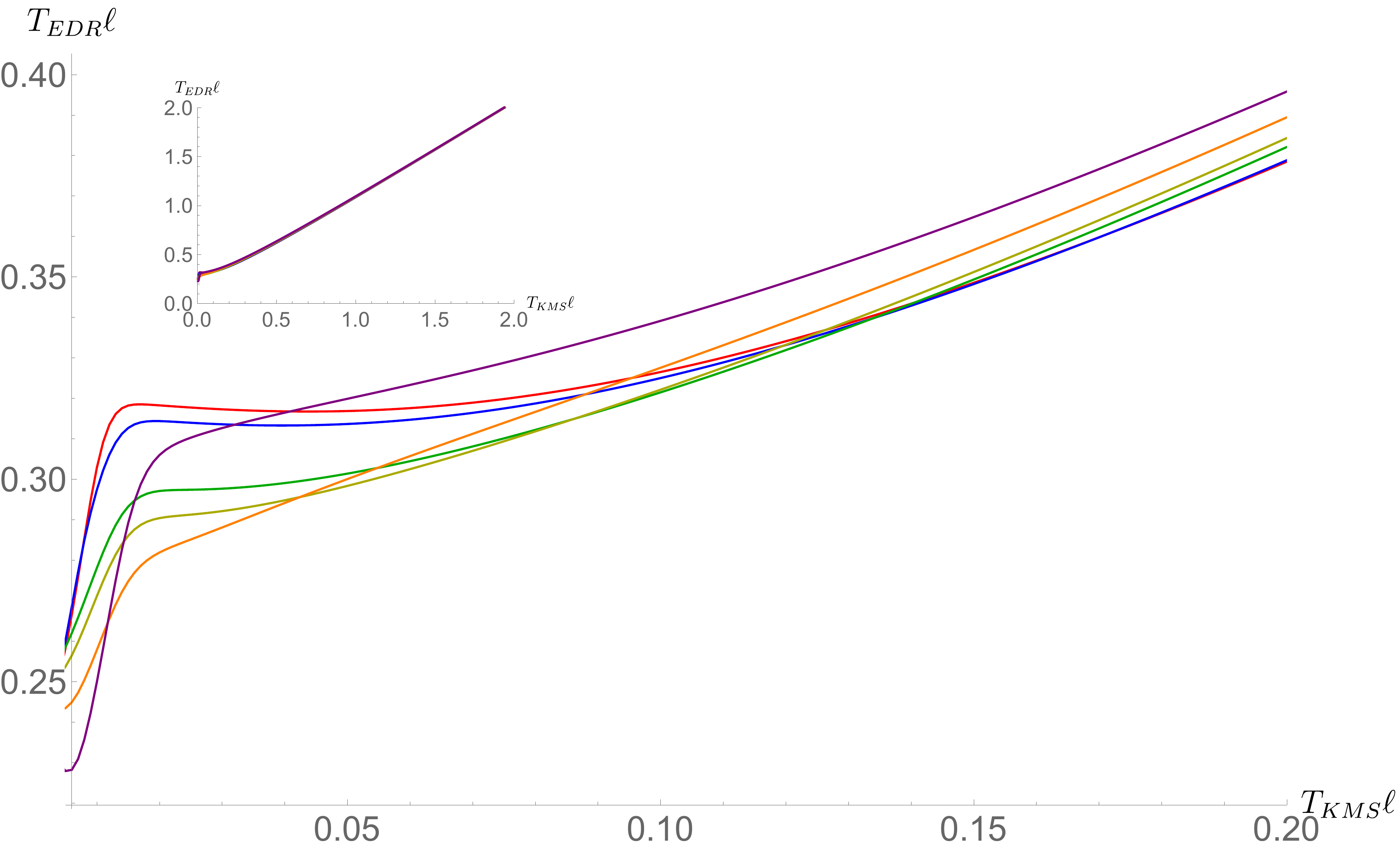}
    \caption*{(a) $\zeta=1$}
\end{minipage}
\begin{minipage}{0.3\textwidth}
  \centering
    \includegraphics[width=\textwidth]{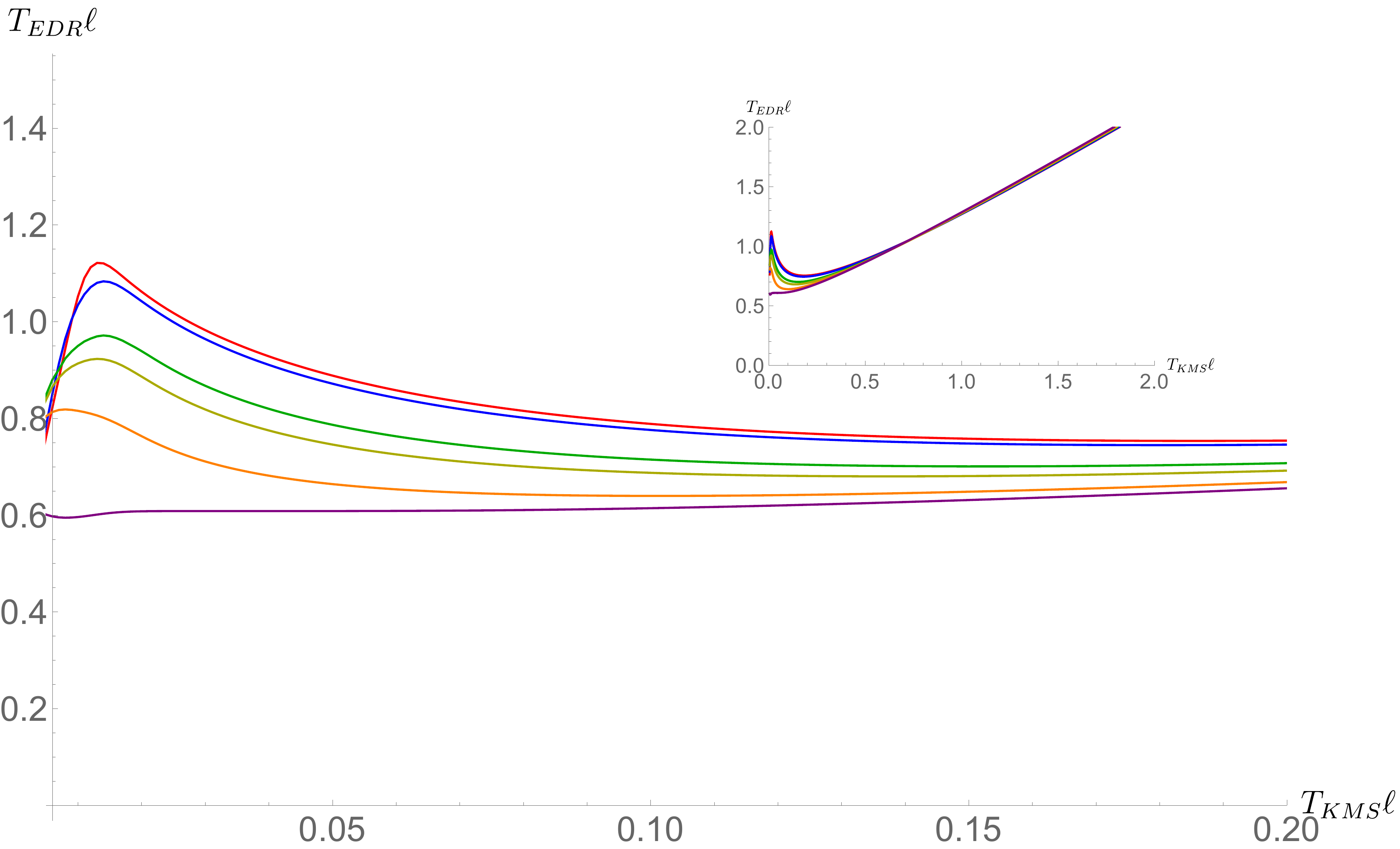}
    \caption*{(b) $\zeta=0$}
\end{minipage}
\begin{minipage}{0.3\textwidth}
  \centering
    \includegraphics[width=\textwidth]{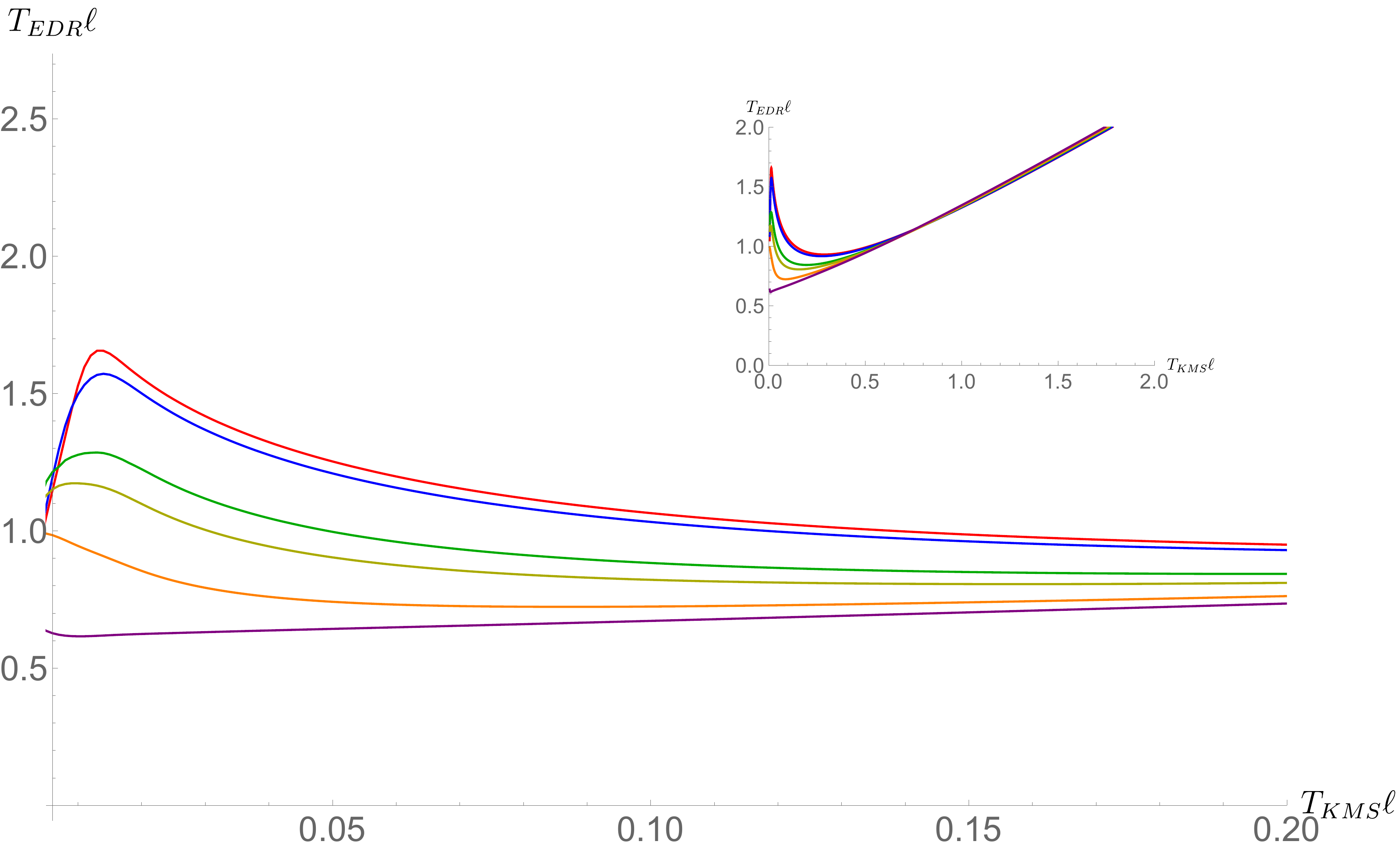}
    \caption*{(c) $\zeta=-1$}
\end{minipage}
\caption{EDR temperature for a black hole of mass $M=1/1000$ and energy gap of $\Omega\sigma=1$. We plot KMS temperature down to $T_{KMS}\ell=10^{-5}$. The insets show the relation between the EDR temperature and KMS temperature for larger values of $T_{KMS}$.} 
\label{fig: TEDRKMS}
\end{figure*}

\begin{figure}
  \centering
    \includegraphics[width=\columnwidth]{PDLegendConstantM.pdf}
    \begin{minipage}{0.45\textwidth}
  \centering
    \includegraphics[width=\columnwidth]{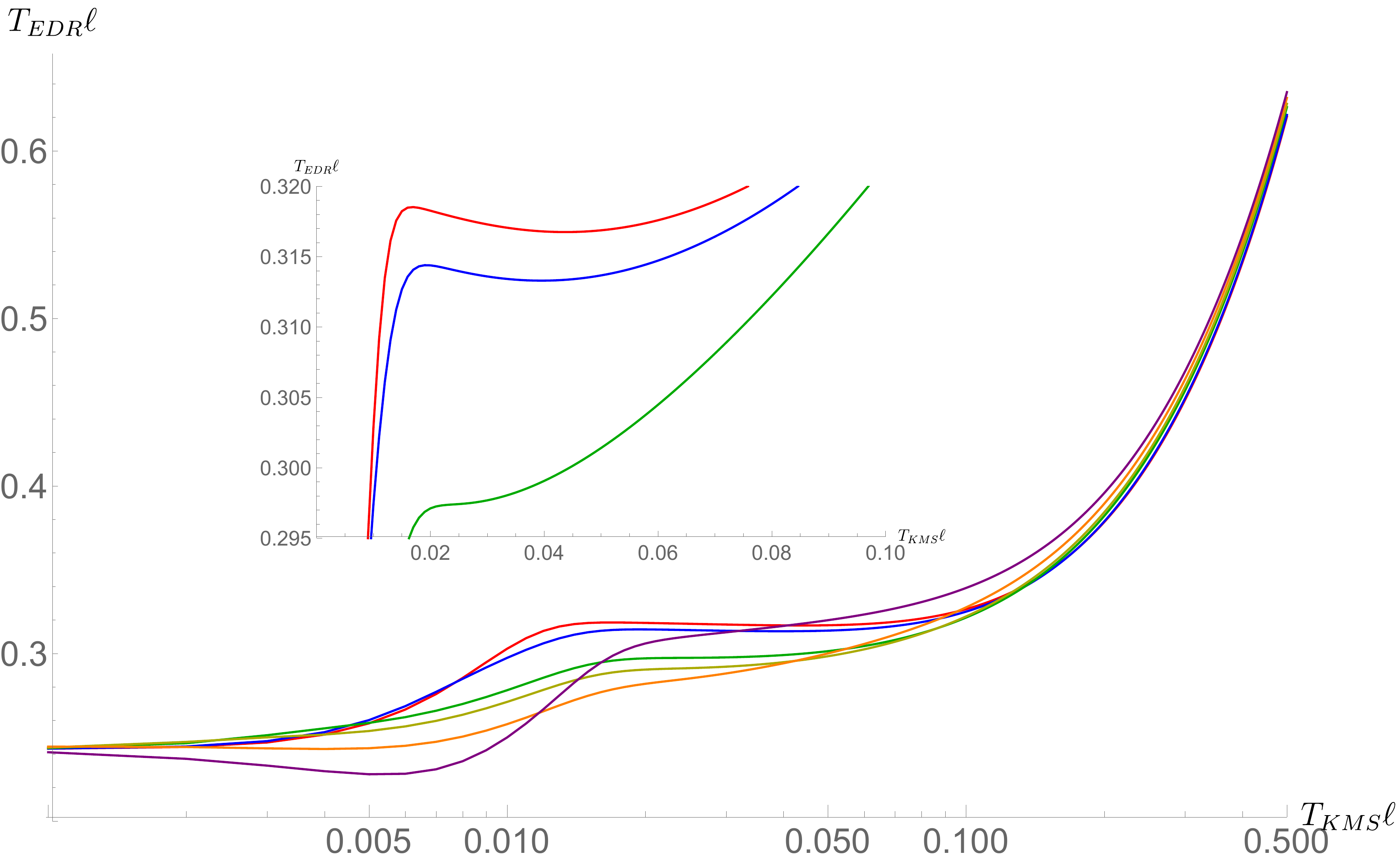}
        \caption*{(a) $\Omega\sigma=1$}
    \end{minipage}
    \begin{minipage}{0.45\textwidth}
  \centering
\includegraphics[width=\columnwidth]{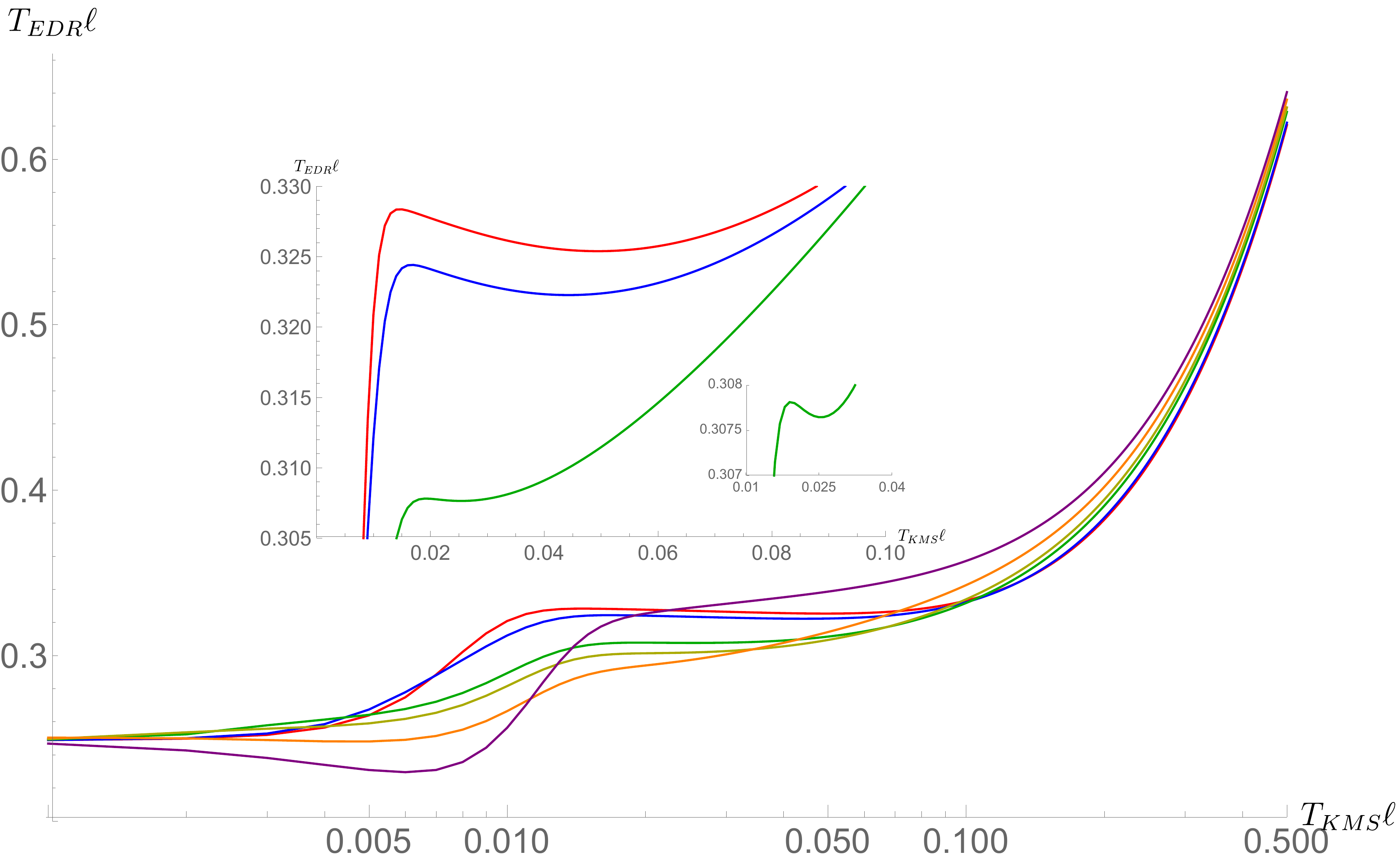}
    \caption*{(a) $\Omega\sigma=1/100$}
\end{minipage}
\caption{EDR temperature for a black hole of mass $M=1/1000$. The results are similar for $\Omega\sigma=1/100$.  The insets show the EDR temperature for our first three values of the angular momentum, plotted on a linear scale (rather than a log scale, as is the case for the main plots).
}
\label{fig: strong anti-Hawking Omega=1/100}
\end{figure}

Let us now turn our attention to the strong anti-Hawking effect. In Figures~\ref{fig: TEDRKMS} and~\ref{fig: strong anti-Hawking Omega=1/100}, we plot the relationship between the EDR   and KMS temperatures for $M=1/1000$ and different boundary conditions, with $\Omega\sigma=1$ and $\Omega\sigma=1/10$, respectively. We note several interesting features.

 First, it is evident from Figure~\ref{fig: TEDRKMS} that a strong anti-Hawking effect is  present for all three boundary conditions. 
This is clear at low values of $T_{KMS}$, where we see a negative slope indicative of the strong effect. Eventually a minimum
is reached and $T_{EDR}$ begins to increase with $T_{KMS}$. 
The insets indicate the behaviour at large $T_{KMS}$, where we see that this quantity is indeed positively correlated with
$T_{EDR}$, and there is minimal dependence on angular momentum, even for small $M$.
 Unlike the weak effect, we see that the maximum decreases with increasing values of angular momentum for all boundary conditions. Consequently the strength of the strong effect (the slope in equation \eqref{strongAH})
 likewise diminishes.
 As extremality is approached, the strong effect essentially vanishes.

\begin{figure}
  \centering
    \includegraphics[width=\columnwidth]{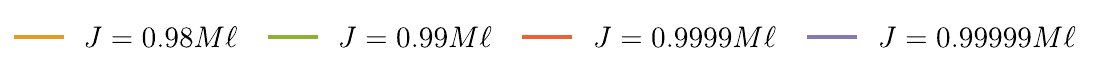}
    \begin{minipage}{0.45\textwidth}
  \centering
    \includegraphics[width=\columnwidth]{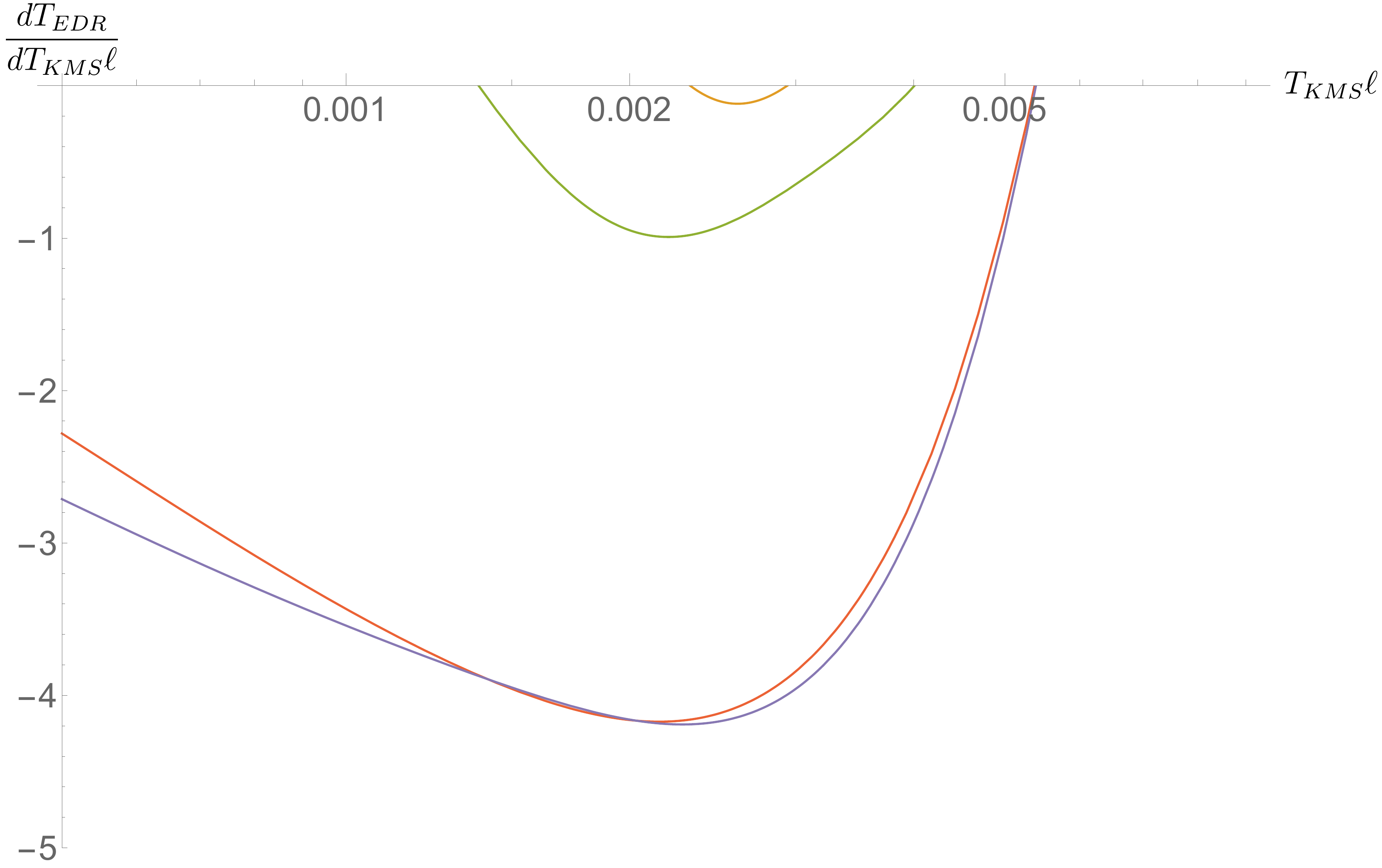}
        \caption*{(a) $\Omega\sigma=1$}
    \end{minipage}
    \begin{minipage}{0.45\textwidth}
  \centering
\includegraphics[width=\columnwidth]{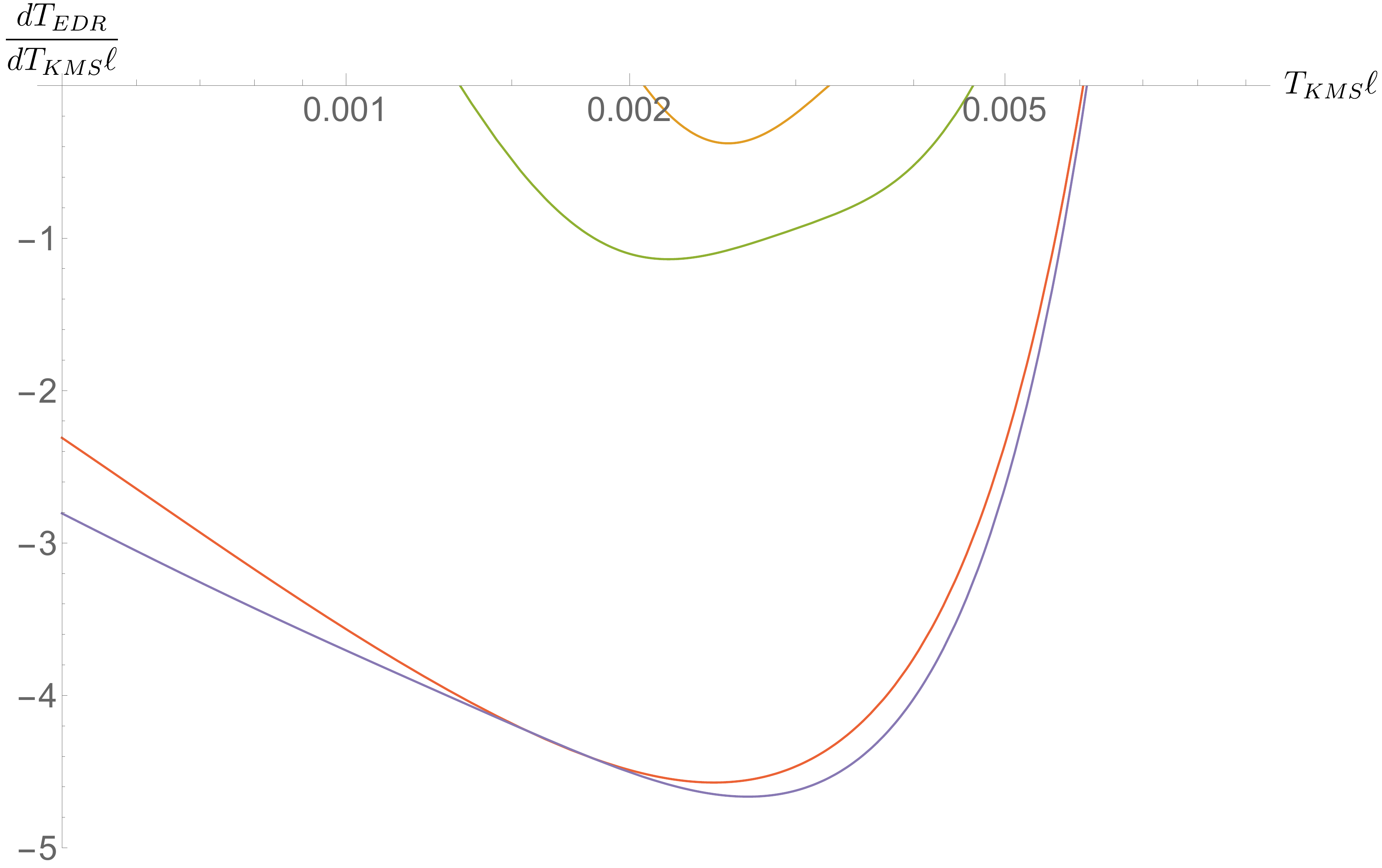}
    \caption*{(a) $\Omega\sigma=1/100$}
\end{minipage}
\caption{ Strong anti-Hawking effect for a near-extremal black hole of mass $M=1/1000$ and Dirchlet boundary conditions. The results are similar for $\Omega\sigma=1/10$.
}
\label{fig: strong anti-Hawking derivative}
\end{figure}

The exception is for Dirichlet boundary conditions.  In Figure~\ref{fig: strong anti-Hawking Omega=1/100}, we plot the EDR temperature for Dirichlet boundary conditions for $\Omega\sigma=1$ and $\Omega\sigma=1/100$. In the insets, we see that there is a small strong anti-Hawking effect for non-rotating black holes, similar to what was found in \cite{Henderson:2019uqo}. As the angular momentum increases,  the peak in Figure~\ref{fig: strong anti-Hawking Omega=1/100} moves rightward, and the threshold value of $T_{KMS}$ at which the strong effect appears increases.   This threshold value
will reach a maximum for some $J/M\ell$, after which
 it decreases with increasing  $J/M\ell$ as extremality is approached, as is clear from Figure~\ref{fig: strong anti-Hawking derivative}.
We also qualitatively see that the magnitude of the slope becomes larger for larger angular momenta, indicating that the strong anti-Hawking effect becomes stronger. For smaller values of the energy gap, we see that the strong anti-Hawking effect increases in strength, with the $\Omega\sigma=1/10$ case being similar to $\Omega\sigma=1/100$.  This can be seen by noting that $J/M\ell=0.9$ does not exhibit the strong anti-Hawking effect for $\Omega\sigma=1$, but there is a small effect present at this value of the angular momentum for $\Omega\sigma=1/100$. 

For transparent and Neumann boundary conditions, we find that the strong anti-Hawking effect is much more pronounced, as can be seen from Figure~\ref{fig: TEDRKMS}.   
Furthermore, as $\zeta$ increases from $-1$ to $1$ in Figure \ref{fig: TEDRKMS}, the range of $T_{KMS} \ell$ over which the strong effect is present also decreases, and is very small for   Dirichlet boundary conditions.
 %We also see that the range of $T_{KMS} \ell$  for which the strong anti-Hawking effect is present 
This range decreases with increasing angular momentum for transparent and Neumann boundary conditions. There is a minimal value of $T_{EDR}\ell$ as function of $T_{KMS} \ell$, and this minimal value decreases as the angular momentum of the black hole increases.
 
  For Dirichlet boundary conditions and sufficiently small angular momentum, larger angular momenta also yields a decreasing range of $T_{KMS}$ temperatures for which the strong effect holds, up to a critical value of $J/M\ell$. Beyond this value, increasing angular momenta results in a greater range of $T_{KMS}$ temperatures  for which the effect is present. As a result, the anti-Hawking effect appears to nearly disappear for near-extremal black holes $J \geq 0.9999 M \ell$ for transparent and Neumann boundary conditions, yet is still present for Dirichlet boundary conditions. In Figure \ref{fig: strong anti-Hawking Omega=1/100}, we see that there is a strong anti-Hawking effect for a non-rotating black hole, but the effect disappears (or almost disappears, depending on the energy gap) as the angular momentum increases to $J/M\ell=0.9$. 
  
Beyond this, however, as we continue to approach extremality in the Dirichlet case, the strong effect emerges at lower KMS temperatures, as shown in   Figure~\ref{fig: strong anti-Hawking derivative}.  Indeed, its range and strength both get larger
as $J/M\ell$ gets very close to unity, as evidenced by the curve for $J/M\ell=0.9999$.   It is quite remarkable that there is such a strong dependence on boundary conditions.   
 
Finally,  inspection of Figures~\ref{fig: TEDRKMS} and \ref{fig: strong anti-Hawking Omega=1/100} indicate that the strong
effect does not monotonically depend on $J/M\ell$.   Indeed, we observe a `crossover' effect at small $T_{KMS}$,  
in which the values of
$T_{EDR}$ decrease with increasing angular momentum as   $T_{KMS} \to 0$, whereas at sufficiently large $T_{KMS}$ the rate of change of  $T_{EDR}$  with respect to $T_{KMS}$ increases with increasing angular momentum such that the higher-$J$ curves cross over the lower-$J$ curves.  At large $T_{KMS}$, we see that $T_{EDR}\sim T_{KMS}$, with the largest $T_{EDR}$ corresponding to the largest $J$ for fixed $T_{KMS}$, and the smallest $T_{EDR}$ corresponding to the smallest $J$.
This is clearly evident in Figure~ \ref{fig: strong anti-Hawking Omega=1/100}. 
We have also verified that this effect is also present for Neumann and transparent boundary conditions, though the `crossover' occurs at larger KMS temperatures than the Dirichlet case.

%. For small $T_{KMS}$, greater angular momentum implies small EDR temperature. However, for sufficiently large values of the KMS temperature, this relation completely reverses, so that the the largest EDR temperature at fixed $T_{KMS}$ corresponds to the largest angular momentum, with the smallest EDR temperature corresponding to the smallest angular momentum.
 
 By comparing Figures \ref{fig: FJ}-\ref{fig: strong anti-Hawking Omega=1/100}, we see that there is no range of $T_{KMS}$ for which
 the strong anti-Hawking effect   overlaps  
 with the weak anti-Hawking effect,  with the weak anti-Hawking effect appearing for $T_{KMS}\ell\gtrsim1$, while the strong anti-Hawking effect appears for $T_{KMS}\ell\lesssim0.1-0.5$. The exact temperature range is dependent on the boundary conditions, energy gap, and in the case of the strong anti-Hawking effect, the angular momentum.
 Furthermore, the critical KMS temperature at which the strong anti-Hawking effect disappears becomes smaller for larger angular momentum, in contrast to the weak anti-Hawking effect where the critical temperature 
at which this  effect appears has minimal dependence on angular momentum.  In addition, we again note that the location of this critical temperature for the strong effect is highly dependent on the boundary conditions.

 %Let us now turn our attention to energy gap. As we saw for the weak anti-Hawking effect, a smaller energy gap decreases the temperature range of that effect as well as amplifies its strength. For the strong anti-Hawking effect, the strength of the effect is similarly amplified  for smaller gap (albeit, not as much as the weak anti-Hawking effect), though there is minimal dependence on the temperature range of that effect. \tcr{\bf [refer to how this is illustrated - which curves are compared - also true for purple?]}

\begin{figure}
  \centering
    \includegraphics[scale=0.5]{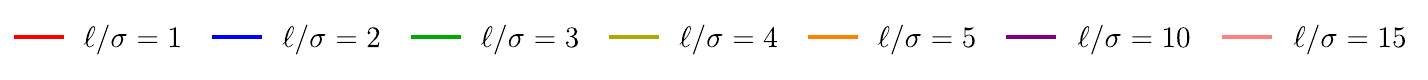}
        \begin{minipage}{0.45\textwidth}
  \centering
\includegraphics[width=\columnwidth]{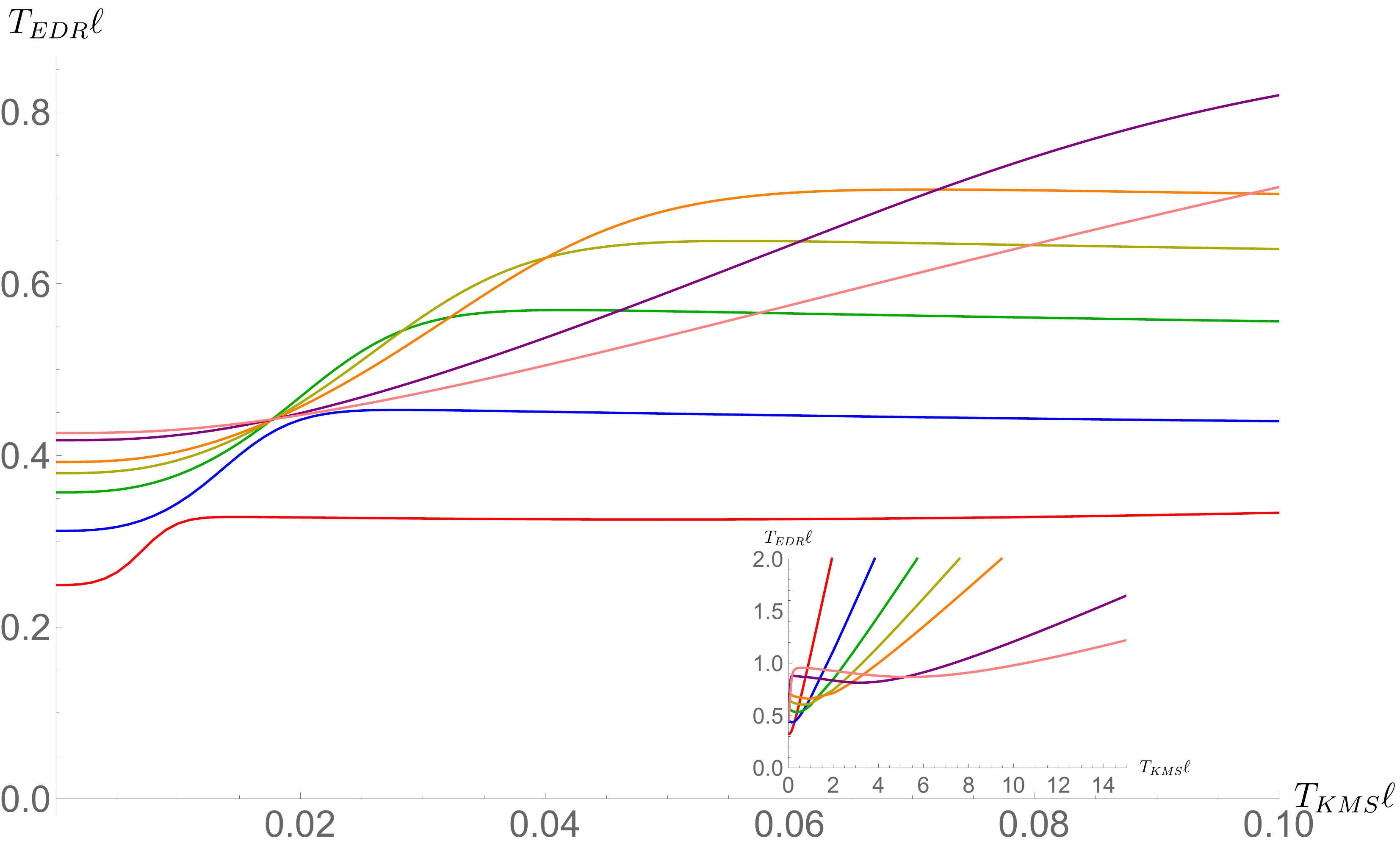}
        \caption*{(a) $J/M\ell=0$}
\end{minipage}
    \begin{minipage}{0.45\textwidth}
  \centering
\includegraphics[width=\columnwidth]{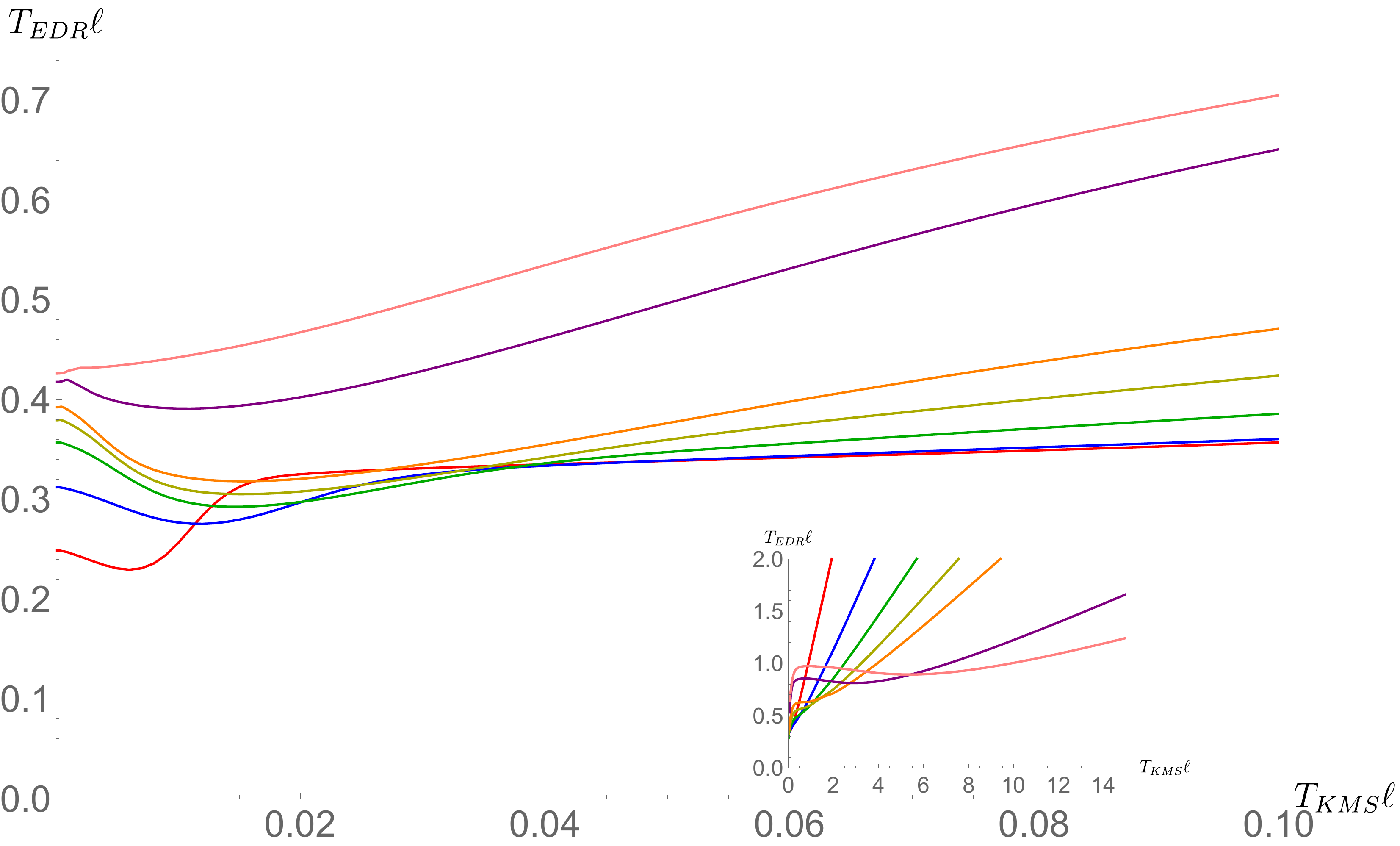}
        \caption*{(b) $J/M\ell=0.9999$}
\end{minipage}
\caption{Changing AdS lengths for the strong anti-Hawking effect for a black hole of mass $M=1/1000$, Dirichlet boundary conditions, and energy gap of $\Omega\sigma=1/10$.  We plot KMS temperature down to $T_{KMS}\ell=10^{-5}$. The insets show the effect of changing AdS length on the EDR temperature for larger values of $T_{KMS}$.
}
\label{fig: strong anti hawking changing AdS length}
\end{figure}

Our last consideration is that of  the impact of changing the AdS length on the strong  effect. Here the situation differs from the
weak anti-Hawking effect  as now there is a second length scale present ($\sigma$, the width of the switching function). In Figure~\ref{fig: strong anti hawking changing AdS length}, we consider the effect of changing the AdS length for a non-rotating BTZ black hole, compared to a near-extremal BTZ black hole. In the  non-rotating case,   increasing AdS length increases the range of $T_{KMS}$ temperatures where the strong anti-Hawking effect holds. However, the marginal effect of increasing $\ell$ is reduced for larger and larger values of the AdS length.
We also see that for small $T_{KMS}\ell$, a larger AdS length will  broaden
 the initial peak. In the case of a near-extremal black hole, the situation is similar. As we saw in Figure \ref{fig: TEDRKMS}a, there was a tiny strong anti-Hawking effect present for near-extremal black holes. For larger AdS lengths, we similarly see that the temperature range of the strong anti-Hawking effect increases in size. However, we note that this effect is still relatively weak and only becomes noticeable for larger values of $\ell$.

\section{Conclusion}

 As with entanglement harvesting  \cite{Robbins2020}, 
rotation can have a significant impact on the anti-Hawking effect. For large-mass black holes with Dirichlet and transparent boundary conditions, the weak anti-Hawking effect vanishes, as expected; it is present for Neumann boundary conditions
\cite{Henderson:2019uqo}. In all cases the effects of rotation are negligible. But as the mass of the black hole decreases, rotation
 significantly amplifies the weak version of the effect.  
 
 The impact of rotation on the strong effect is somewhat inverted.  We find that rotation tends to weaken the strength of the strong anti-Hawking effect for transparent and Neumann boundary conditions, with it nearly vanishing for near-extremal black holes. In contrast, for Dirichlet boundary conditions, larger angular momenta causes the strong anti-Hawking effect to be reduced before being amplified again.

Furthermore, for the strong anti-Hawking effect, the relationship between angular momentum and detector temperature is
non-monotonic  for each boundary condition, leading to a `crossover' phenomenon that is most prominent for Dirichlet boundary conditions. For small $T_{KMS}L$, larger angular momenta yield smaller $T_{EDR}L$, whereas for larger values of  $T_{KMS}L$,   larger angular momenta yield \textit{larger}  $T_{EDR}$.  More work is needed to better understand how this cross-over effect comes about and its dependence on the  boundary conditions. 
It would also be interesting to consider whether $3+1$ dimensional rotating black holes also exhibit these same findings as the rotating BTZ black hole. 
 
While the weak anti-Hawking effect is independent of  the AdS length $\ell$, 
increasing AdS length increases the range of $T_{KMS}$ temperatures where the strong anti-Hawking effect holds.   
A larger AdS length will also broaden the initial peak  for small $T_{KMS}\ell$.
However as  $\ell$ continues to increase, its impact on the strong effect becomes increasingly marginal. 

In summary, our results indicate that the effects of spacetime dragging on the quantum vacuum can significantly modify
detector response as small field KMS field temperatures, as exemplified by the anti-Hawking effect(s). The role of boundary
conditions is very important; indeed, it is surprising that there is such a strong dependence of the strong anti-Hawking effect on boundary conditions for  small-mass rotating black holes.  The origin of this effect merits further study. 
\bigskip

{\it Acknowledgements}
$\quad$ 
MR was funded by an Ontario Graduate Scholarship. This research was supported in part by the Natural Sciences and Engineering Research Council of Canada, Asian Office of Aerospace Research \& Development Grant FA2386-19-1-4077,  and the Perimeter Institute for Theoretical Physics. Research at Perimeter Institute is supported in part by the Government of Canada through the Department of Innovation, Science and Economic Development Canada and by the Province of Ontario through the Ministry of Colleges and Universities.

\bibliographystyle{unsrt}
\bibliography{AntiHawkingBibliography}

\end{document}